\documentclass[aps,amssymb,12pt,showpacs,showkeys,letterpaper]{revtex4}
\usepackage{graphicx}
\usepackage{amsmath}
\usepackage{epsfig}
\usepackage{bm}

\newcommand{\be}{\begin{equation}}
\newcommand{\ee}{\end{equation}}
\newcommand{\beq}{\begin{eqnarray}}
\newcommand{\eeq}{\end{eqnarray}}

\begin{document}

\title{Harmonic Oscillator in Snyder Space, the Classic and the Quantum}

\author{Carlos Leiva }
 \email{cleivas@uta.cl}
\affiliation{\it Departamento de F\'{\i}sica, Universidad de
Tarapac\'{a}, Casilla 7-D, Arica, Chile}

\date{\today}

\begin{abstract}
The harmonic oscillator in Snyder space is investigated in its
classical and quantum versions. The classical trajectory is
obtained and  the semiclassical quantization from the phase space
trajectories is discussed. In the meanwhile, an effective cutoff
to high frequencies is found. The quantum version is developed and
an equivalent usual harmonic oscillator is obtained through  an
effective mass and  an effective frequency introduced in the
model. This modified parameters give us an also modified energy
spectra.
\end{abstract}



\maketitle
\section{\label{sec:Int}Introduction}
Today, the possibility that the space could be noncommutative is
an idea more and more accepted. The non-commutativity is usually
set through a constant parameter \cite{Hor,Poli,gamb}.But exists
another more general formulation, that is the Snyder space
\cite{Sny}. R. Snyder  investigated these ideas long time ago and
built a noncommutative Lorentz invariant space-time where the
non-commutativity of space operators is proportional to non-linear
combinations of phase space operators through a free parameter
$l$, that is usually identified with the Planck longitude
$l_p=\sqrt{G/c\hbar}$ . Kontsevich \cite{Kon}, worked out these
kind of space and since then, Snyder-like spaces in the sense of
non-commutativity are of ever-increasing interest. Snyder space is
also interesting because it can be mapped to the k-Minkowski
space-time \cite{KG}. This space can be canonically and elegantly
obtained in its  classical version through a lagrangian and
hamiltonian approach \cite{rabin}, and a dimensional reduction
from a (D+1,2) space with two time dimensions, to a $(D,1)$ space
with just one time dimension \cite{zam0}.

Nowadays Snyder space is also  of increasing interest because it
could be seen as an environment where it could be possible being
successfully in quantizing gravity. In fact, in that direction it
is possible to find a plausible explanation to the Bekenstein
conjecture for the area spectrum of a black hole horizon through
the area quantum in this kind of space \cite{zam1}.

In this paper the harmonic oscillator is analyzed in its classical
and quantum versions. The quantum version of this simple but
fundamental system was studied in \cite{34}, but the treatment
here is  simpler   and we don't need to bet about the right
operators and despite that, we can shed some light in probably
applications to problems like infinities in Quantum Field Theory
(QFT). That is the importance of building a well defined harmonic
oscillator in this kind of space, it could be possible to develop
a QFT in it with very desirable properties. Furthermore, in the
paper it is shown that we can build an Harmonic Oscillator with an
effective mass related to the $l$ parameter.

The paper is organized as follows. In section $2$, the classical
version is investigated and some possible quantum consequences
postulated, in section $3$ the quantum version is developed and
the energy spectra is obtained. Finally the results are discussed
in section $4$

\section{The Classic}
Classical $n$ dimensional  Snyder Space is characterized by its
non linear commutation relation between the variables of the phase
space:

\begin{eqnarray}
\{q_i,q_j\}&=&-l^2L_{ij}, \\ \label{cnc}
\{q_i,p_j\}&=&\delta_{ij}-l^2p_ip_j,\\ \label{cnl}
\{p_i,p_j\}&=&0,
\end{eqnarray}
where $l$ is a tiny constant parameter (usually identified with
Planck longitude),  that measures the deformation introduced in
the canonical Poisson brackets, and $L_{ij}$, the angular
momentum.

 Let's consider the
usual Hamiltonian of an Harmonic Oscillator :

\begin{equation}
H=\frac{1}{2}p^2+\frac{\omega^2}{2}q^2,
\end{equation}

where $m=1$, so $\omega^2=k$. The canonical equations are:

\begin{equation}
\dot{q}=\{q,H\}=p-l^2p^3 \label{qder},
\end{equation}

and for p:
\begin{equation}
\dot{p}=\{p,H\}=-\omega^2q+\omega^2l^2qp^2. \label{ppunto}
\end{equation}

If we solve $p(q)$, we find the usual relation
$p^2+\omega^2q^2=\omega^2$. So the orbits in the phase space are
untouched after the deformation of the Poisson brackets.

Solving the simultaneous equations (\ref{qder}) and (\ref{ppunto})
we obtain for q:

\begin{equation}
q=\pm \frac{\tan\{(\omega
t+d)\sqrt{1-l^2\omega^2}\}}{\sqrt{\frac{1}{1-l^2\omega^2}+\tan^2\{(\omega
t+d)\sqrt{1-l^2\omega^2}\}}},
\end{equation}

where $d$ is a suitable constant in order to achieve the initial
condition $q(t=0)=1$,  and p can be expressed in terms of q:

\begin{equation}
p=\pm \omega \sqrt{1-q^2}.
\end{equation}

\begin{figure}[!h]
  \begin{center}
    \includegraphics[width=80mm]{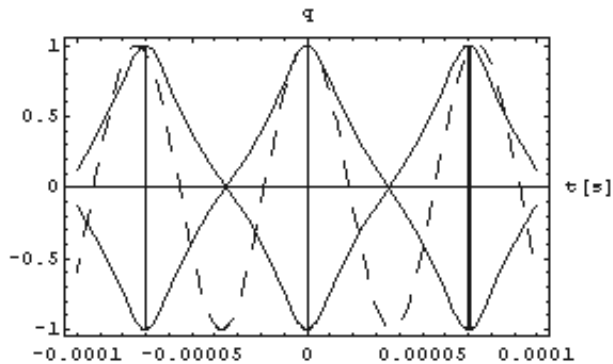}
  \end{center}
  \caption{Plot  for the two branches of Snyder $q(t)$ with $l=10^{-5}$ and $\omega=8.5\cdot 10^{-4}$(continue line), and for normal $q(t)$ (dashed).}
  \label{fig:g1}
\end{figure}
\begin{figure}
  \begin{center}
    \includegraphics[width=80mm]{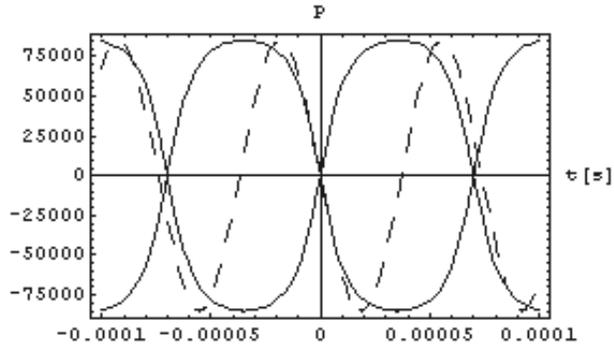}
  \end{center}
  \caption{Plot  for the two branches of Snyder $p(t)$ with $l=10^{-5}$ and $\omega=8.5\cdot 10^{-4}$(continue line), and for normal $p(t)$ (dashed).}
  \label{fig:g1}
\end{figure}

Figures $1$ and $2$ show the behavior of the positions $q$ and
momentum $p$. We can see from $q$ graph, that Snyder oscillator is
periodic but contains  harmonics that deform the trajectory.
Furthermore, it is possible to see that the Snyder oscillator has
a different equivalent period than the normal one.

Another conspicuous feature is that $\omega=1/l$ is effectively a
cutoff to high frequencies. Indeed this is a good new in the
search of possible  QFT theory in Snyder space, because there is
some hope of avoiding infinities.

Due to the orbits are not affected by the non linear version of
Poisson brackets, we expect that the energy spectra from the
Sommerfeld-Wilson quantization method $\int pdq=n\hbar$, should be
formally the same that the one of the linear oscillator:
$E_n=n\hbar w$, but as we have seen, the real equivalent period is
not the same as the one of the normal oscillator, so  some
variations in the values should be expected.

Of course, Snyder Oscillator is not longer an expression in single
$Sin$ or $Cos$ functions i.e it is not a pure oscillator, but we
still can express it as a Fourier transform and formulate it as a
linear infinite combination of harmonics of the frequency
$\omega$. That is, the single  Snyder oscillator looks like a set
of normal coupled oscillators.

In fact, using the fact that $l^2$ must be considered as a tiny
parameter in relation with the all other quantities, we can use a
perturbation method, among others,  to solve these equations.

Let's start with the usual solution to $p$:
\begin{equation}
p_0=-\omega\sin(\omega t),
\end{equation}
where we normalized the initial $q$ perturbation, that is:
$q(t=0)=1$. With $p_0$, we can integrate $\dot{q}$ in order to
obtain the first order $q_1$:

\begin{equation}
q_1(t)=(1-\frac{3}{4}l^2 \omega^2)\cos(\omega
t)+\frac{1}{12}l^2\omega^2\cos(3\omega t)+ q_1^0,
\end{equation}

where $q_1^0$ is the constant evaluated in order of having the
initial value of $q$. Now, we can introduce $q_1$ in
(\ref{ppunto}) and integrate to obtain $p_1$

\begin{equation}
p_1(t)=(-1+l^2 \omega^2-\frac{5}{24}l^4\omega^5)\sin(\omega
t)+(-\frac{1}{9}l^2
\omega^3+\frac{11}{144}l^4\omega^5)\sin(3\omega
t)-\frac{1}{240}l^4\omega^5\sin(5\omega t),
\end{equation}

and so on. The method gives us, as we expected, the expansion in
terms of harmonic functions in harmonic frequencies of $\omega$.

\section{The Quantum}

After the Dirac quantization recipe, we can postulate the
commutation relations of the Snyder Space:
\begin{equation}
[\hat{Q}_i,\hat{Q}_j]=-il^2\hat{L}_{ij}, \label{nc}
\end{equation}
\begin{equation}
[\hat{Q}_i,\hat{P}_j]=i\delta_{ij}-il^2\hat{P}_i\hat{P}_j,
\label{nl}
\end{equation}
\begin{equation}
[\hat{P}_i,\hat{P}_j]=0.
\end{equation}

The (\ref{nc}) relation is a nonlinear version of the usual non
commutativity  one, where  the commutator of the position
operators is proportional to a constant \cite{Poli,gamb}. Here it
is proportional to  the angular momentum operator:  $\hat{L_{ij}}$

The (\ref{nl}) equation is related to the different models with
generalized commutation relations \cite{18,33,yo}

\bigskip

 To study the one dimensional Harmonic Oscillator we start considering an
standard Hamiltonian:
\begin{equation}
\hat{H}=\frac{1}{2m}\hat{P}^2+\frac{1}{2}m\omega^2\hat{X}^2.
\end{equation}

We define the usual creation and annihilation operators, where
$\hbar=1$:

\begin{eqnarray}
a=\sqrt{\frac{m\omega}{2}}\hat{Q}+i\sqrt{\frac{1}{2m\omega}}\hat{P},\\
a^\dag=\sqrt{\frac{m\omega}{2}}\hat{Q}-i\sqrt{\frac{1}{2m\omega}}\hat{P}.
\end{eqnarray}

Using the commutation rules between operators of position and
momentum, the commutation rules of the operators $a$ and $a^\dag$
are $[a,a]=[a^\dag,a^\dag]=0$, and:

\begin{equation}
[a,a^\dag]=1-l^2\hat{P}^2=1+\frac{l^2}{2}(a^\dag-a)^2.
\end{equation}

Writing the Hamiltonian in terms of  the creation and annihilation
operators and using the commutation relation between $a$ and
$a^\dag$, we obtain:

\begin{equation}
H= \omega \{a^\dag a+\frac{1}{2})+ \frac{\omega l^2}{2}(a^\dag
a^\dag-a^\dag a- a a^\dag+ a^2\}.
\end{equation}

Due the structure of the Hamiltonian, $|n\rangle$ is not longer an
eigenvalue of the Hamiltonian, in fact:

\begin{eqnarray}
H|n\rangle= &\omega &
\{n[1-\frac{l^2}{(1+l^2)}]+\frac{1}{2}[1+\frac{l^2}{(1+l^2)}]\}|n\rangle
\nonumber
\\
+&\omega & \{\frac{l^2}{2(1+l^2)}\sqrt{n+1}\sqrt{n+2}\}|n+2\rangle
\nonumber
\\
+&\omega& \{\frac{l^2}{2(1+l^2)}\sqrt{n}\sqrt{n-1}\}|n-2\rangle.
\end{eqnarray}
So, the Snyder oscillator mixes states as we expected from the
classical version.

But, encouraged by the semiclassical quantization that says us
that we could find an standard spectra of the energy, we will use
a QFT trick: because the extra term in the Hamiltonian induced by
the non linearity of the commutators of $a$ and $a^\dag$ is
proportional to the dynamic term, we can add an counter term to
the original Hamiltonian:
\begin{equation}
\widetilde{H}=\frac{1}{2m}\hat{P}^2+\frac{1}{2}m\omega^2\hat{X}^2+\frac{l^2}{2}\hat{P}^2
\end{equation}

Now, it is possible to define  a new mass   parameter, $
\widetilde{m}=m/(1+ml^2)$ and modify the frequency,
$\widetilde{\omega}=\omega\sqrt{(1+ml^2)}$, then $\widetilde{H}$
becomes:

\begin{equation}
\widetilde{H}=\frac{1}{2\widetilde{m}}\hat{P}^2+\frac{1}{2}\widetilde{m}\widetilde{\omega}^2\hat{X}^2.
\end{equation}

This Hamiltonian  has as eigenvector $|n-2\rangle$ and as
eigenvalue $n$, and its spectra is:

\begin{equation}
E=\widetilde{\omega}(n+\frac{1}{2}).
\end{equation}

This mass renormalization like procedure allows us to see the
Snyder oscillator as the usual one, at least in the energy
spectra, but with an effective mass. The energy spectra have been
modified due the $l$ parameter. So, the zero energy is
$\widetilde{E}_0=\widetilde{\omega}/2$ and the $\triangle
E=\widetilde{\omega}$ (remember that $\hbar=1$).

\section{Discussion and Outlook}
In the paper we have found the classical trajectory of an
oscillator in Snyder space and found that we can see it as a set
of coupled oscillators that can be described by an expansion in
harmonic functions in the harmonic frequencies of $\omega$. We
found also that there is a high frequency cutoff, because beyond
it the oscillator has no response. Due this, we can hope that
infinities in QFT theories could be avoided in Snyder space.
Furthermore, we could see that the Sommerfeld-Wilson quantization
method indicates us that the spectra should be formally like the
usual harmonic oscillator, and consequentially in the quantum
formulation, we saw that the oscillator in this space effectively
mixes states, but through a QFT of mass renormalization, we could
build an standard harmonic oscillator with an energy spectra
modified due the presence of the non-commutative parameter $l$. We
could expect that in the following we will can couple infinite
oscillators in order to built an effective QFT in Snyder space. On
the other hand, it will be worth to investigate the movement
integrals of this kind of systems in higher dimensions.

\begin{acknowledgments}
C.Leiva was supported by Grant UTA DIPOG N$^0$ 4721-07
\end{acknowledgments}

\end{document}